\documentclass[aps,prc,groupedaddress,showpacs,twocolumn,floatfix]{revtex4}
\usepackage[mathscr]{eucal}
\usepackage[german,english]{babel}
\usepackage[intlimits,sumlimits,namelimits,reqno]{amsmath}
\usepackage[dvips]{epsfig}
\usepackage{float}
\usepackage{amssymb,ifthen,shadow}
\usepackage{fancyhdr}
\usepackage{colordvi}

\begin{document}

\title{Correlations in Hot Asymmetric Nuclear Matter}
\author{T. Frick, H. M\"uther,}
\affiliation{Institut f\"ur Theoretische Physik, \\
Universit\"at T\"ubingen, D-72076 T\"ubingen, Germany}
\author{A. Rios, A. Polls, and A. Ramos}
\affiliation{Departament d'Estructura i Constituents de la Mat\`eria,\\
Universitat de Barcelona, E-08028 Barcelona, Spain}

\begin{abstract}
The single-particle  spectral functions in asymmetric nuclear matter are 
computed using the ladder approximation within the theory of finite temperature
Green's functions. The   internal energy and the momentum distributions of
protons and neutrons are studied as a function of  the density and the
asymmetry of the system. The  proton states are more  strongly depleted when
the asymmetry increases while the occupation of the neutron states is enhanced
as compared to the  symmetric case. The self-consistent Green's function
approach leads to slightly smaller energies as compared to the Brueckner
Hartree Fock approach. This effect increases with density and thereby modifies
the saturation density and leads to smaller symmetry energies. 

\end{abstract}
\maketitle

\section{Introduction}
The equation of state (EOS) of asymmetric nuclear matter is a necessary
ingredient in the description of astrophysical environments of interest such as
supernova explosions  or the structure of neutron stars \cite{mor1}. Actually,
the study of asymmetric nuclear matter is also relevant to understand stable
nuclei because they themselves are asymmetric  nuclear systems with a different
number of protons and neutrons. The recent availability of data concerning
nuclei far form the stability valley has also renovated the interest in the 
study of asymmetric nuclear matter as a first step in the microscopic study of
these nuclei.  

The evaluation of this EOS starting from realistic models of the
nucleon-nucleon (NN) interaction is still one of the challenging open problems
in nuclear physics. In fact, the presence of strong-short range and tensor
components in the realistic NN interactions, which are required to fit the NN
scattering data, are the origin of the  corresponding correlations in the
nuclear wave function. The study of these correlations  and their influence on
different observables has recently  made an important progress not only from
the theoretical side   but also from the experimental point of
view \cite{bal1,her1,dick1}.
 In the later  case, the analysis of $(e,e'p)$ reactions on
$^{208}$Pb covering a wide range of missing energies lead to the conclusion
that the occupation numbers for the deeply bound proton states are depleted by
about 15-20 \% \cite{exp1}. This depletion can be identified with the
corresponding depletion  of hole states in nuclear matter with momenta well
below the Fermi momentum \cite{ram1,fabro1,her2}.

Several theoretical tools have been developed and applied to describe these
correlations in nuclear systems. These include the Brueckner hole-line
expansion \cite{day} and also variational  approaches using correlated basis
functions \cite{akmal1,fabro2,fan1}.

Recently, an enormous progress has been achieved in using the self-consistent
evaluation of Green's function \cite{kad62,kraeft} to solve the nuclear
many-body problem \cite{her1,dick1,dick2,fri03,die1,boz2,boz3,boz4}. This
method gives direct access to the single particle spectral function,  i.e.
to the single particle properties and, in particular, to the occupation
numbers. At the same time, the modification of the single particle properties
affects   the effective interactions between nucleons in the medium and both
things should be determined in a self-consistent way.  

Most of the microscopic calculations have been addressed to study symmetric
nuclear matter (SNM) and  pure neutron matter (PNM) \cite{fri03,die1}. The study
of asymmetric nuclear matter is technically more involved  and only few
Brueckner-Hartree Fock calculations are available \cite{bombaci,isa1,hassan}. 
In most of the cases, one assumes a quadratic dependence of the energy per
particle 
\begin{equation}
\label{asy_energy}
\frac{E}{A}(\rho,\alpha) = \frac {E}{A}(\rho,0) + a_s(\rho) \alpha^2 +\dots\,.
\end{equation}
in terms of the asymmetry  parameter, $\alpha = (N-Z)/A$  and the symmetry
energy $a_s(\rho)$. In this way,  the calculation of the energy of SNM and PNM
allows to  determine the symmetry energy $a_s(\rho)$ and using the previous
equation  one can estimate the energy for any asymmetry.  That this quadratic
expression is a good approximation has been directly confirmed  in
Brueckner-Hartree-Fock calculations of asymmetric nuclear
matter \cite{bombaci,isa1}.  In the case of the variational approach,
calculations  for asymmetric matter with the same accuracy that the ones
performed for SNM or PNM are not yet available. However, besides the energy per
particle which is governed mainly by the symmetry energy, there are other
observables, as the momentum distributions, which do not necessary follow a
quadratic dependence. 

In this paper we want to perform a calculation of asymmetric nuclear matter in
the framework of SCGF theory. The calculation is performed at finite
temperature. A temperature of 5 MeV has been chosen because it is small 
enough to allow for conclusions for the $T=0$ case. On the other hand this
temperature is large enough to allow for a smooth numerical representation of
the spectral functions and to avoid the possibility of proton-neutron pairing 
instabilities \cite{von1,alm,boz1}.   In the next section, we briefly describe
some specific features of the SCGF formalism for asymmetric nuclear matter. 
The discussion of the results and a systematic comparison with results of the
Brueckner-Hartree-Fock approach is presented in Section III. The main
conclusions are summarized in the last section.

\section {Formalism} 
A key quantity in the theory of Green's functions --- that allows to compute
all single-particle observables as well as the internal energy of the system
--- is the spectral function $A(k,\omega)$. It can be obtained as a solution of
Dysons's equation,
\begin{equation}
\label{spec_function}
A_{\tau}(k,\omega)=\frac{-2\,{\mathrm{Im}}\,\Sigma_{\tau}(k,\omega+{\mathrm{i}}
\eta)}{[\omega-\frac{k^2}{2m}-{\mathrm{Re}}\,\Sigma_{\tau}(k,\omega)]^2+
[{\mathrm{Im}}\,\Sigma_{\tau}(k,\omega+{\mathrm{i}}\eta)]^2}.
\end{equation}
where $\Sigma_{\tau}(k,\omega+{\mathrm{i}}\eta)$ is the retarded self
energy of a nucleon with isospin projection $\tau$, which can
either be a proton ($\tau=+\frac{1}{2}$ or $p$) or a neutron
($\tau=-\frac{1}{2}$ or $n$).

In the ladder or $T$~matrix approximation, the self energy contains an energy
independent Hartree-Fock part and a complex dispersive contribution that
accounts for correlations between the particles. The Hartree-Fock contribution
to the ladder self energy of a nucleon $\tau$ involves an explicit sum over the
isospin projection $\tau^{\prime}$ of the internal particle,
\begin{eqnarray}
\label{eq_hf_partialwaves} \Sigma^{HF}_{\tau}(k)& = &
\frac{1}{4\pi}
\sum_{\tau^{\prime}}\!\!\! \sum_{\substack{JSL\\
T\le|\tau+\tau^{\prime}|}}\!\!\! (2J+1)\,\,|C^{T
\tau+\tau^{\prime}}_{\frac{1}{2}\tau \frac{1}{2} \tau^{\prime}}|^2\nonumber \\
&&\quad\times\int \frac{{\mathrm{d}}^3k^{\prime}}{(2\pi)^3}
\left<q\right|V^{JST}_{LL}\left|q\right>
n_{\tau^{\prime}}(k^{\prime}).
\end{eqnarray}
with $V^{JST}_{LL^{\prime}}$ the nuclear two-body potential in a
partial wave representation. The relative momentum between the
interacting particles is given by
$\mathbf{q}=\frac{1}{2}(\mathbf{k}-\mathbf{k}^{\prime})$ and the
Clebsch-Gordan coefficient for the isospin quantum numbers is denoted by
$C^{T
\tau+\tau^{\prime}}_{\frac{1}{2}\tau \frac{1}{2} \tau^{\prime}}$.
The momentum distribution,
\begin{equation}
\label{occupation} n_{\tau}(k)= \int_{-\infty}^{+\infty}
\frac{{\mathrm{d}}\omega}{2\pi}
A_{\tau}(k,\omega)f_{\tau}(\omega),
\end{equation}
must be derived from the non-trivial spectral function. The Fermi-Dirac
function for nucleons with isospin projection $\tau$ is denoted by
$f_{\tau}(\omega)=\{\exp[\beta(\omega-\mu_{\tau})]+1\}^{-1}$, where $\beta$
stands for the inverse temperature $T^{-1}$. 
For a given total density $\rho$, the
partial fraction $x_{\tau}=\rho_{\tau}/\rho$ of the respective
particle species is given by
\begin{equation}
\label{eq_density} x_{\tau}=\frac{\gamma}{\rho}
\int\frac{{\mathrm{d}}^3k}{(2\pi)^3} n_{\tau}(k),
\end{equation}
where $\gamma=2$ is the spin degeneracy factor of the system. The
partial fractions of protons and neutrons add up to one,
$x_p+x_n=1$ and the asymmetry is given by
$\alpha = x_n - x_p\,$. 
Considering a fixed composition, Eq.~(\ref{eq_density}) can be used to fix the
chemical potential $\mu_{\tau}$.

The generalized expression for the imaginary part of the self
energy in the $T$~matrix approximation for the case of asymmetric
nuclear matter reads:
\begin{eqnarray}
\label{eq_ladder_partialwaves}
\lefteqn{{\mathrm{Im}}\Sigma_{\tau}(k,\omega+{\mathrm{i}}\eta) =}\\
& &\frac{1}{4\pi}
\sum_{\tau^{\prime}}\!\!\! \sum_{\substack{JSL\\
T\le|\tau+\tau^{\prime}|}}\!\!\! (2J+1)\,\,|C^{T
\tau+\tau^{\prime}}_{\frac{1}{2}\tau \frac{1}{2} \tau^{\prime}}|^2
\int\frac{{\mathrm{d}}^3k^{\prime}}{(2\pi)^3}
\int_{-\infty}^{+\infty} \frac{{\mathrm{d}}\omega^{\prime}}{2\pi}\nonumber\\
&&A_{\tau^{\prime}}(k^{\prime},\omega^{\prime})
 \left<q\right|\!
{\mathrm{Im}}T^{JST\,\tau+\tau^{\prime}}_{LL}\!(P,\omega+\omega^{\prime}+
{\mathrm{i}}\eta)\!\left|q\right>\nonumber\\
&&\,\times
[f_{\tau^{\prime}}(\omega^{\prime})+b_{\tau,\tau^{\prime}}
(\omega+\omega^{\prime})].\nonumber
\end{eqnarray}
Here,
$b_{\tau,\tau^{\prime}}(\Omega)=\{\exp[\beta(\Omega-\mu_{\tau}-
\mu_{\tau^{\prime}})]-1\}^{-1}$
is the Bose distribution function. The total pair momentum is
given by $P=\frac{1}{2}|\mathbf{k}+\mathbf{k}^{\prime}|$.
A dispersion relation that is reported, e.g., in Ref.~\cite{fri03}
determines the real part of  the dispersive contribution to the ladder 
self energy.
The $T$~matrix elements contain the re-summation of the ladder
diagrams to all orders and can be obtained as the solution of a
scattering-type integral equation,
\begin{eqnarray}
\label{eq_twaves}\lefteqn{
\left<q\right|T^{JST\,m_T}_{LL^{\prime}}(P,\Omega+{\mathrm{i}}\eta)
\left|q^{\prime}\right> =}\\
&& \left<q|V^{JST}_{LL^{\prime}}|q^{\prime}\right>
 + \sum_{L^{\prime \prime}} \int_0^{\infty} \frac{ {\mathrm{d}}
k^{\prime}\,k^{\prime 2} } {(2\pi)^3}
\left<q|V^{JST}_{LL^{\prime \prime}}|k^{\prime}\right>\nonumber \\ &&
\times {\bar{g}}_{m_T}^{\mathrm{II}}(P,\Omega+{\mathrm{i}}\eta,k^{\prime})
 \left<k^{\prime}\right|T^{JST\,m_T}_{L^{\prime \prime}L^{\prime}}
(P,\Omega+{\mathrm{i}}\eta)\left|q^{\prime}\right>.\nonumber
\end{eqnarray}
Since in ANM, the in-medium propagation of a neutron is different
from that of a proton, the $T$~matrix elements depend upon the
third component of the isospin of the propagating pair,
$m_T=\tau+\tau^{\prime}$, via the non-interacting two-particle
propagator ${g}_{m_T}^{\mathrm{II}}$,
\begin{eqnarray}
\label{two_pp}\lefteqn{
g_{\tau_1+\tau_2}^{\mathrm{II}}(k_1,k_2,\Omega+{\mathrm{i}}\eta)=}\\
&&\nonumber\int_{-\infty}^{+\infty}\frac{{\mathrm{d}}\omega}{2\pi}
\int_{-\infty}^{+\infty}\frac{{\mathrm{d}}\omega^{\prime}}{2\pi}
A_{\tau_1}(k_1,\omega)A_{\tau_2}(k_2,\omega^{\prime})\\
&&\nonumber \times
\frac{1-f_{\tau_1}(\omega)-f_{\tau_2}(\omega^{\prime})}
{\Omega-\omega-\omega^{\prime}+{\mathrm{i}}\eta}.
\end{eqnarray}
To circumvent the coupling between partial waves of different
total angular momenta, ${g}_{m_T}^{\mathrm{II}}$ enters
Eq.~(\ref{eq_twaves}) in an angle-averaged form that is indicated
by the bar.

Eqs.~(\ref{spec_function})--(\ref{two_pp}) must be solved
self-consistently for a given temperature $\beta^{-1}$, a given
density $\rho$ and a given proton fraction $x_p$. Partial waves up
to $J=9$ were included for the calculation of the
energy-independent part of the self energy. The dispersive part
includes only partial waves up to $J=2$. A more comprehensive
description of the numerical details is given in
Ref~\cite{fri03}. The numerical routine that was developed
for ANM was tested in the following way: the results for SNM and
PNM are recovered if one chooses $x_p=0.5$ and $x_p=0$,
respectively. Furthermore, we have checked that charge symmetry is
fulfilled, which can be expressed by the condition
$\Sigma_p(k,\omega,x_p)=\Sigma_n(k,\omega,1-x_p)$.  \\
Once the self-consistent solution for the spectral function has
been obtained, the internal energy per particle in the system is
given by
\begin{equation}
\frac{E}{A}=\frac{\gamma}{\rho}\sum_{\tau}
\int\frac{{\mathrm{d}}^3k^{\prime}}{(2\pi)^3}
\int_{-\infty}^{+\infty} \frac{{\mathrm{d}}\omega^{\prime}}{2\pi}
\frac{1}{2}\left(\frac{k^2}{2m}+\omega\right)
A_{\tau}(k,\omega)f_{\tau}(\omega).
\end{equation}

The BHF approach is obtained from the previous formulation by assuming that
the  single particle spectral functions are characterized by only one energy
having the full  strength accumulated in this energy, $A_{\tau} (k, \omega) =
\delta(\omega - \epsilon_{\tau}^{BHF}(k))$, being $\epsilon_{\tau}^{BHF}(k)$
the BHF single particle energy.  In addition, in the  two-body propagator
(Eq.(\ref{two_pp})) one only considers the propagation of particle states.
Notice also that the BHF self-energy does not include the contribution of the
Bose distribution that  appears in Eq.(\ref{eq_ladder_partialwaves}). 

\section{Results and Discussion}

All the results discussed in this paper have been  computed for the
charge-dependent Bonn (CDBONN) potential, defined in \cite{mach1}
which is non-local and exhibits a softer tensor
component compared to other  realistic potentials as the Argonne V18
\cite{arg1} or Reid93 \cite{reid93}, which are local.  
Since we want to concentrate on the asymmetry dependence, we
will only consider one temperature $T=5$ MeV, low enough  for the conclusions on
the asymmetry dependence to be valid at $T=0$ MeV and high enough  to avoid the
instabilities associated to the neutron-proton pairing. 

\begin{figure}[htb]
\begin{center}
\epsfig{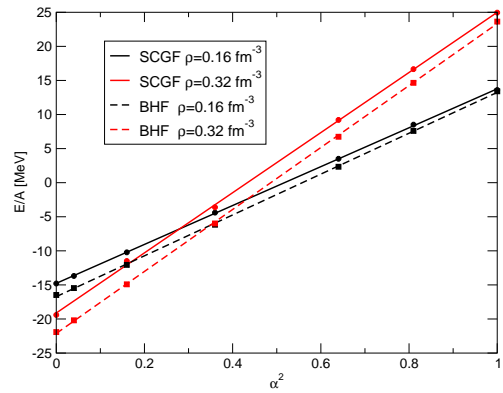}
\end{center}
\caption{\label{fig.fig1} Dependence of E/A on $\alpha^2$ computed in the SCGF
approach (full lines) and in the BHF approach (dashed lines) at two densities,
$\rho=0.16 $ fm$^{-3}$(black lines) and $\rho=0.32 $ fm$^{-3}$(red lines).}
\end{figure}

The binding energies of asymmetric nuclear matter calculated in the BHF
approximation and in the framework of the SCGF   theory are reported in Fig.
(\ref{fig.fig1}) as a function of the square of the asymmetry
parameter $\alpha^2$.  The plots correspond to two densities,  the empirical
saturation density of symmetric nuclear matter, $\rho=0.16$ fm$^{-3}$, and
twice this density.  Notice, however, that BHF calculations using the
CDBONN potential yield a  saturation point at higher densities.  

The first thing to realize is the linear dependence of the energy in terms of
$\alpha^2$,  for both types of calculations (BHF and SCGF) in the full range of
variation of the asymmetry from  SNM to PNM. In the case of BHF, this fact has
already been  considered in the literature \cite{bombaci} and provides the
justification to use PNM and SNM to  determine the coefficients in  Eq.
(\ref{asy_energy}), which turns out to be a very good approximation.  

The propagation of holes in the SCGF and the consideration of the spectral
functions in the intermediate states of the ladder equation result in a
repulsive effect with respect to the continuous choice BHF calculation. This
repulsive effect  increases with density, being 1.8 MeV  and 3.2 MeV in SNM
matter at $\rho=0.16$ fm$^{-3}$ and $\rho=0.32$ fm$^{-3}$, respectively. 

However, for a given density, the difference between the SCGF and the BHF 
calculation does not depend so much on the asymmetry,  being slightly larger in
SNM than in PNM. This leads to a small decrease  of the symmetry energy of the
SCGF calculation as compared to the BHF one. At $\rho=0.16$ fm$^{-3}$,  $a_s=
30.0$ MeV in the BHF approximation while the SCGF schemes provide $a_s =28.6$
MeV. This is in qualitative agreement with a recent comparison of  both
approaches \cite{die1}, in which the symmetry energy has been deduced from SNM
and PNM calculations. The reduction of the symmetry obtained in the present
analysis, however, is weaker than the decrease by around 4 MeV reported in
\cite{die1}.  In this reference,  however, the Reid93 potential,  which has a
stronger tensor component and also stronger short range correlations than the 
CDBONN, was employed. Furthermore it should be noted that the spectral
functions employed in \cite{die1} are described in terms of three $\delta$
functions, whereas a continous description has been used in the present
approach. In addition, the calculation of \cite{die1} was performed at $T=0$
whereas we consider $T=5$ MeV. We do not believe, however, that the temperature 
can be the origin of this discrepancy. In fact, the effect of the finite
temperature is a slight increase of the symmetry energy, but this will also be 
true in a  similar amount for the BHF approach. Instead this discrepancy
demonstrates that the loss of energy in SCGF as compared to BHF is a delicate
balance between an increase of kinetic energy and more attractive potential 
energy. In fact, the perturbative inclusion of the hole-hole scattering terms
of \cite{hassan} even leads to an increase of the symmetry energy.

\begin{figure}
\begin{center}
\epsfig{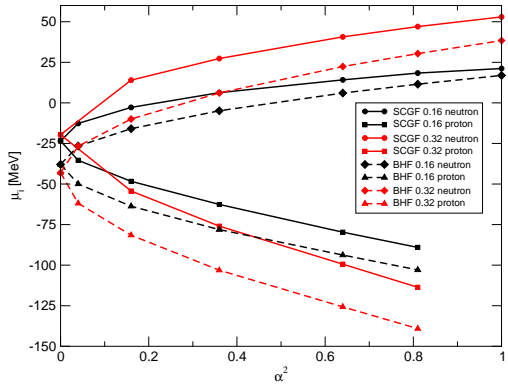}
\end{center}
\caption{\label{fig.fig2} Dependence of neutron (upper curves) and proton
(lower curves) chemical potentials $\mu$ on the asymmetry $\alpha$ calculated
in the SCGF approach (solid lines) and on the BHF approach (dashed curves) at
two given densities, $\rho=0.16 $ fm$^{-3}$(black lines) and $\rho=0.32 $
fm$^{-3}$(red lines).}
\end{figure}

The second point that we want to consider is the dependence of the chemical
potential  of protons and neutrons on the asymmetry. Fig. (\ref{fig.fig2})
shows the chemical potential of neutrons  (upper curves) and protons (lower
curves) calculated in the SCGF approach  (solid lines) and in the BHF approach
(dashed curves).  The chemical potential of protons and neutrons coincide at
the symmetric case ($\alpha=0$), as we have not considered the
charge symmetry breaking  terms contained in the CDBONN potential. When  the
asymmetry increases, the neutron chemical potential increases and becomes
positive, while the one of  the protons becomes more and more attractive. 
The dependence on $\alpha^2$ is not linear anymore.  One should keep
in mind that in the case of SCGF, the chemical potential that one obtains from
the normalization condition of the partial density should  coincide with the
one obtained from the free energy using the thermodynamic relation
$\mu_{\tau}=F/A+\rho\frac{\partial F/A}{\partial \rho_{\tau}} $.  On the
contrary, it is well known that in the BHF approach the chemical potential
derived from the normalization condition substantially differs from the one 
obtained by the thermodynamic relation. Actually, this is already true at $T=0$
MeV where one needs to incorporate the rearrangement terms in the self-energy
in order to recover the relation $\mu = \epsilon(k_f)$ (where
$\epsilon(k_F)$ is the  single-particle energy at the Fermi surface). 

It is important to note that the differences between the BHF and the SCGF
chemical potentials are much larger (of about $\sim 15$ MeV in the whole range of
asymmetries) than the differences in the energy per particle discussed in Fig.
(\ref{fig.fig1}), thus indicating that the role of correlations has a larger
influence in this observable than in the energy per particle and also giving an
idea of the magnitude of the rearrangement term in such calculations.
In
particular we want to emphasize that the difference between the chemical
potentials for neutrons and protons is larger in the case of BHF as compared to
SCGF. This implies that the SCGF tends to predict $\beta$-equilibrium with a
smaller proton fraction as derived from BHF calculations.

\begin{figure}
\begin{center}
\epsfig{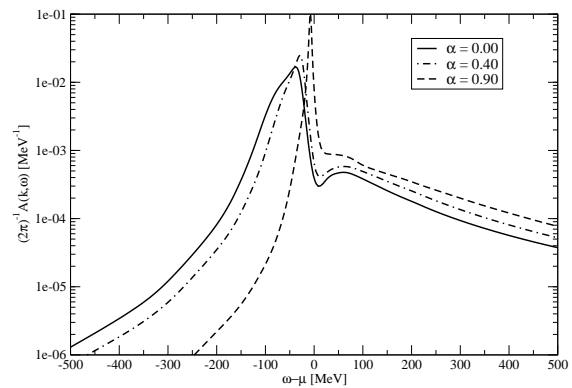}
\end{center}
\caption{\label{fig.fig3} $k=0$ MeV proton spectral function for different
proton fractions. The calculations were performed at $\rho=0.16 $ fm$^{-3}$and
$T=5$ MeV.}
\end{figure}

The next point we shall address is the discussion of the single-particle
spectral  functions. Fig. (\ref{fig.fig3}) shows the proton spectral functions
at $\rho=0.16 $ fm$^{-3}$ for the momentum $k=0$ MeV at different proton
fractions as a function of the energy measured with respect to the proton
chemical potential corresponding  to each fraction. As the asymmetry increases,
the amount of protons decreases and the Fermi-momentum for the protons 
gets closer to $k=0$ MeV. As a consequence, the coupling to two-hole one-particle
configurations with this momentum of $k=0$ is reduced, the quasi-particle peak 
obtains a smaller width and the peak gets higher. The spectral function
at positive energies, however, is larger with increasing asymmetry. This can be
explained by the fact that  correlation effects mainly originate from the tensor
force, which is more important in the proton-neutron than in the neutron-neutron
or proton-proton interaction. Note that the most important partial wave for
these tensor correlations is the $^3S_1-^3D_1$ channel, which is relevant for
nucleon pairs with isospin zero only. Since the density of neutrons increases
with the asymmetry parameter, the protons display stronger correlation effects
at these larger asymmetries leading to an enhancement for the spectral function
for $k<k_F$ at energies $\omega >\mu$.  

\begin{figure}
\begin{center}
\epsfig{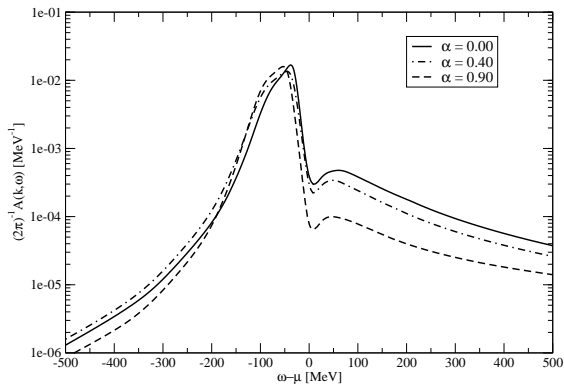}
\end{center}
\caption{\label{fig.fig4} $k=0$ MeV neutron spectral function for different
proton fractions. The calculations were performed at $\rho=0.16 $ fm$^{-3}$ and
$T=5$ MeV.}
\end{figure}

The corresponding plot for the neutron spectral functions  is presented  in
Fig. (\ref{fig.fig4}). Again we consider the momentum $k=0$, the same density
$\rho=0.16 $ fm$^{-3}$ and asymmetries. Looking at the spectral functions at
energies above the chemical potential for the neutrons, we can see that the
correlations which are responsible for the spectral function in this regime are
reduced with the asymmetry, i.e. with the density of protons. This can again be
understood from the dominance of the proton-neutron interaction leading to
these correlations. The width of the neutron spectral functions seems not to be
affected very much by an increase of the asymmetry for the considered values of
$\alpha$. This is a result of the fact that the damping of the strong isospin
zero correlations at larger values of $\alpha$ is counterbalanced by an
increasing phase space for the neutron-neutron configurations. Similar
observations have also been made in the perturbative calculations of asymmetric
matter in \cite{hassan}.

\begin{figure}
\begin{center}
\epsfig{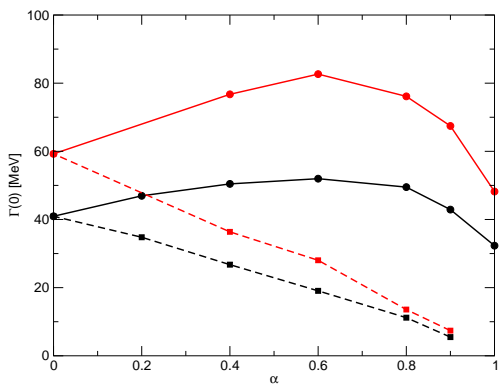}
\end{center}
\caption{\label{fig.fig5} Dependence of the $k=0$ MeV on-shell width
$\Gamma(0)$ on the proton fraction for neutrons (full lines) and protons
(dashed lines) at two given densities, $\rho=0.16 $ fm$^{-3}$(black lines) and
$\rho=0.32 $ fm$^{-3}$(red lines).}
\end{figure}

In order to explore a bit more the width and height of the quasi-particle peak
in the spectral function and its dependence on the asymmetry   we show in Fig.
(\ref{fig.fig5}) the  width of the spectral function  given by $ 2 \mid Im
\Sigma (k, \epsilon_{qp}) \mid $, where $\epsilon_{qp}$ is  the quasi-particle
energy. Obviously, for the symmetric case both the widths for protons and
neutrons coincide. When the asymmetry increases, the width of the protons
decreases monotonically, indicating that $k=0$ is getting closer and closer to
the  proton Fermi surface.  The situation for the neutrons is a little bit
different.  For low asymmetries, the width initially increases with the
asymmetry, reaches a maximum and then it decreases again - being a little lower
for pure neutron matter than for the symmetric case.  This indicates that there
is an asymmetry for which the correlation  effects on the width of the neutron
quasi particle peak reaches a maximum.  As mentioned above, there is a 
competition between the correlations originated by the propagation of
neutron-neutron holes, which should increase when the asymmetry gets larger as
the associated phase space also grows,  and the correlations originating from
the interaction of the neutrons with the protons. It is clear that these last 
correlations get less important for the neutrons with increasing asymmetry.  
Finally, for neutron matter, the decrease of the number of protons and
therefore the suppression of the tensor components of the neutron-proton 
interaction  dominates and the width for neutron matter is slightly smaller
than for symmetric nuclear matter.

\begin{figure}
\begin{center}
\epsfig{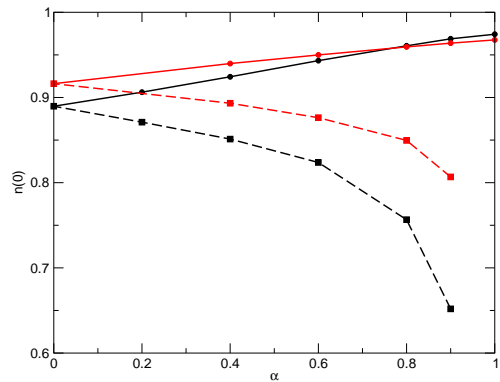}
\end{center}
\caption{\label{fig.fig6} Dependence of the $n(0)$ on the proton fraction for
neutrons (full lines) and protons (dashed lines) at two given densities,
$\rho=0.16 $ fm$^{-3}$(black lines) and $\rho=0.32 $ fm$^{-3}$(red lines).}
\end{figure}

Another observable which has been object of discussion during the last years,
due to its relevance in the analysis of the $(e,e'p)$ reaction is the momentum
distribution, which also provides a clear measurement of the effects 
of 
correlations. In Fig. (\ref{fig.fig6}), we show the occupation of the zero
momentum state as a function  of the asymmetry.  In the symmetric case, we can
observe an unexpected behavior on the occupation as a function of density: at
the larger density there is a higher occupation while at the lower one the
occupation is smaller, a behavior that has been pointed  out previously in the
literature \cite{her2}.  For a given density, the proton depletion   increases
with the asymmetry, indicating the importance of the neutron-proton
correlations. On the  contrary, the depletion of the neutrons gets smaller,i.e
 neutron matter is a less correlated system. Also it is worth to notice that for pure
neutron matter one recovers the expected behavior of  having a higher
occupation for smaller density. 

\begin{figure}
\begin{center}
\epsfig{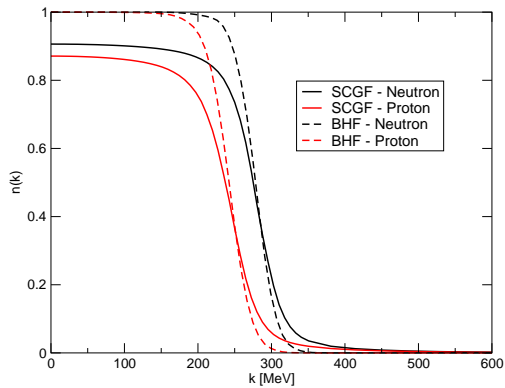}
\end{center}
\caption{\label{fig.fig7} Momentum distributions for neutrons and protons for
the SCGF approach (full lines) and the BHF approach (dashed lines) at a density
$\rho=0.16$ fm$^{-3}$ and asymmetry $\alpha=0.2$. }
\end{figure}

Finally, in Fig. (\ref{fig.fig7}), we show the momentum distribution at
$\rho=0.16$ fm$^{-3}$ and an asymmetry $\alpha =0.2$, which is characteristic
of heavy nuclei, such as $^{208}$Pb. Included in the figure are also the BHF
momentum distributions.  The first thing to notice is that the BHF momentum
distributions do not contain correlation effects and are very similar to a
normal thermal  Fermi distribution. The momentum distributions obtained in the
framework of the SCGF contain, besides the thermal effects, important
short-range and tensor correlations which are reflected in the depletion of
the  occupation at low momentum and in a larger occupation than the BHF
momentum distribution at large  momenta. Notice also that the proton momentum
distribution is more depleted  than the neutron momentum distribution. This is
in agreement with all the previous discussions,  reflecting the fact that the
protons (i.e.the less abundant particle in the realistic asymmetric
conditions)  are more affected by correlations, mainly due to its interactions
with neutrons. This may be  important in the interpretation of the momentum
distributions obtained in the $(e,e'p)$ experiments. Some of the analysis have
been conducted in the framework of  a local density approximation starting from
results obtained in SNM, where the momentum distribution of protons and
neutrons are  identical.  This different behavior of the momentum distributions
of protons and neutrons is in contrast with  the  very recent calculations   by
Bo\.zek \cite{boz4} obtained also in the SCGF framework and with the CDBONN
potential in which no  noticeable difference between the momentum distributions
of neutrons and protons  was found. 

\section {Summary and Conclusions} 

The techniques to evaluate the single-particle Green's function in a
self-consistent $T$-matrix approach (SCGF), that has recently been developed
for nuclear matter \cite{fri03}, have been extended and applied to asymmetric
matter. Actual calculations have been performed using the realistic CDBONN
interaction for asymmetric matter at two densities (the saturation density of
symmetric nuclear matter and twice this density), various asymmetries and for a
temperature of $T=5$ MeV. This temperature is low enough  to allow for 
conclusions on the $T=0$ limit and high enough  to avoid the instabilities
associated to the neutron-proton pairing.

The inclusion of the hole-hole ladders and the self-consistent treatment of the
Green's Function in the SCGF approach leads to a small reduction of the binding
energy per nucleon as compared to the BHF approximation. This effect increases
with density and is slightly weaker for pure neutron matter as compared to
symmetric nuclear matter which leads to a small reduction in the symmetry
energy.

Larger effects are observed for the single-particle properties like the
chemical potential. In particular we observe  in neutron rich matter a
reduction in  the difference between the chemical potential for protons and
neutrons, which  would correspond to smaller proton fractions in $\beta$ stable
matter than those predicted by a  BHF-type equation of state.

The SCGF calculation also yields detailed informations on the single-particle
spectral functions and momentum distributions for protons and neutrons. We
observe a depletion of the proton occupancies for momenta below the
Fermi momentum, which increases significantly with the neutron fraction. This
can be explained by the strong correlations induced from proton neutron
interaction.

\section {Acknowledgments }

Useful discussions with Dr. Isaac Vida\~na are gratefully acknowledged. This
work has been supported by the Europ\"aische  Graduiertenkolleg
T\"ubingen-Basel (DFG-SNF), the German-Spanish exchange program (DAAD, Aciones
Integradas Hispano-Alemanas), the DGICYT (spain) Project No. BFM2002-01868 and
by  the Generalitat de Catalunya Project No. 2001SGR00064.

\end{document}